\documentclass[12pt,a4paper]{article}
\usepackage[utf8]{inputenc}
\usepackage[plainpages=false,pdfpagelabels=true]{hyperref}
\usepackage{amssymb,amsthm}
\usepackage[margin=1.5 in]{geometry}
\usepackage{graphicx}
\usepackage{a4wide}  
\usepackage{amsfonts}
\usepackage{amsmath}
\usepackage{bbm}
\usepackage{booktabs} 
\usepackage{multicol}
\usepackage{ifpdf}
\ifpdf

\hypersetup{%
  pdftitle    = {Light rings, timelike circular orbits and curvature of traversable wormholes}
  pdfkeywords = {wormholes, massive particle surfaces, Jacobi metric, light rings, Timelike circular orbits},
  pdfauthor   = {Oscar},
  plainpages  = true,
  colorlinks  = true,
  citecolor   = blue,
  urlcolor    = blue,
  linkcolor   = black
}

\newcommand{\arxiv}[1]{{\tt
\href{http://www.arXiv.org/abs/#1}{#1}}}
\newcommand{\doi}[1]{{\tt DOI:\href{https://doi.org/#1}{#1}}}

\makeatletter
\@addtoreset{equation}{section}
\makeatother

\begin{document}

\begin{center}

{\Large {\bf Light rings, timelike circular orbits and curvature of traversable wormholes }}

\vspace{1.5cm}

\renewcommand{\thefootnote}{\alph{footnote}}
{\sl\large  Boris Berm\'{u}dez-C\'{a}rdenas\footnote{E-mail: bxbermudez [at] uc.cl }\textsuperscript{b} and Oscar Lasso Andino}\footnote{E-mail: {oscar.lasso [at] udla.edu.ec}}

\setcounter{footnote}{0}
\renewcommand{\thefootnote}{\arabic{footnote}}

\vspace{1.5cm}

{\it $^a$ Facultad de Matemáticas, Pontificia Universidad Católica de Chile, Avenida Vicuña Mackenna 4860, Santiago, Chile.}
\ \vspace{0.3cm}

{\it $^b$Escuela de Ciencias Físicas y Matemáticas,Universidad de Las Américas, Redondel del ciclista, Antigua vía a Nayón, C.P. 170504, Quito, Ecuador\\} \vspace{0.3cm}

\vspace{1.8cm}

%%%%%%%%%%%%%%%%%%%%%%%%%%%%%%%%%%%%%%%%%%%%%%%%%%%%%%%%%%%%%%%%%%%%%%

{\bf Abstract}

\end{center}

\begin{quotation}
We study the existence of light rings (LR's) and timelike circular orbits (TCO's) in spherically symmetric, asymptotically flat wormhole geometries. We use a purely geometric approach based in the intrinsic curvatures of a $2$-dimensional Riemannian metric obtained by projecting the spacetime metric over surfaces of constant energy. Using the asymptotic and near throat limits of geodesic curvature we determine the existence of LR's and TCO's in wormhole geometries, then analyzing the sign of the Gaussian curvature we are able to determine their stability. We deduce the conditions for the existence of photon and massive particle surfaces.  We apply the results to the Morris-Thorne family of wormholes, Damour-Solodukhin wormholes and Ellis-Bronnikov wormholes.  We show that for the wormholes considered there is always an odd number of light rings, one of them at the throat. In the case of Morris-Thorne wormhole our method leads to a procedure for building a wormhole with more than one LR. We study how the procedure can be extended to other wormhole spacetimes.
\end{quotation}

%%%%%%%%%%%%%%%%%%%%%%%%%%%%%%%%%%%%%%%%%%%%%%%%%%%%%%%%%%%%%%%%%%%%%%
%%%%%%%%%%%%%%%%%%%%%%%%%%%%%%%%%%%%%%%%%%%%%%%%%%%%%%%%%%%%%%%%%%%%%%
%%%%%%%%%%%%%%%%%%%%%%%%%%%%%%%%%%%%%%%%%%%%%%%%%%%%%%%%%%%%%%%%%%%%%%
%%%%%%%%%%%%%%%%%%%%%%%%%%%%%%%%%%%%%%%%%%%%%%%%%%%%%%%%%%%%%%%%%%%%%%
\pagestyle{plain}
%%%%%%%%%%%%%%%%%%%%%%%%%%%%%%%%%%%%%%%%%%%%%%%%%%%%%%%%%%%%%%%%%%%%%%
%%%%%%%%%%%%%%%%%%%%%%%%%%%%%%%%%%%%%%%%%%%%%%%%%%%%%%%%%%%%%%%%%%%%%%
%%%%%%%%%%%%%%%%%%%%%%%%%%%%%%%%%%%%%%%%%%%%%%%%%%%%%%%%%%%%%%%%%%%%%%
%%%%%%%%%%%%%%%%%%%%%%%%%%%%%%%%%%%%%%%%%%%%%%%%%%%%%%%%%%%%%%%%%%%%%%
%%%%%%%%%%%%%%%%%%%%%%%%%%%%%%%%%%%%%%%%%%%%%%%%%%%%%%%%%%%%%%%%%%%%%%

%\tableofcontents

\newpage

%%%%%%%%%%%%%%%%%%%%%%%%%%%%%%%%%%%%%%%%%%%%%%%%%%%%%%%%%%%%%%%%%%%%%%
%%%%%%%%%%%%%%%%%%%%%%%%%%%%%%%%%%%%%%%%%%%%%%%%%%%%%%%%%%%%%%%%%%%%%%
%%%%%%%%%%%%%%%%%%%%%%%%%%%%%%%%%%%%%%%%%%%%%%%%%%%%%%%%%%%%%%%%%%%%%%
%%%%%%%%%%%%%%%%%%%%%%%%%%%%%%%%%%%%%%%%%%%%%%%%%%%%%%%%%%%%%%%%%%%%%%

\section{Introduction}
In 2019 one of the most spectacular discoveries was announced. An intriguing object, predicted long time ago,  was detected. A black hole having a mass equivalent to millions of solar masses was perfectly measured by the Horizon Event telescope collaboration \cite{EventHorizonTelescope:2022wkp,Tsukamoto:2023rzd}. These objects have an event horizon that captures everything, including light. Therefore, it is impossible to detect an event horizon directly. The only possible way to capture any information is indirectly through the phenomena that occurs around it. In particular, the vibrations of the light ring (LR) that surrounds the black hole shadow can be measured, and information about the mass and the spin of the black hole can be extracted. However, there are objects that can mimic the black hole LR and therefore they can be confused with black holes \cite{DeSimone:2025sgu,DeFalco:2021btn,DeFalco:2023kqy}. This shows that LR's are not carrying the whole information needed and cannot be sufficient for determining the observed object. \\
The shadow of a wormhole  can be measured in a similar way of that of black hole. However, since a wormhole does not posses a horizon, the shadow of it will stops light rays coming from the other side of the throat. It could happen that a LR can be located only at one side of the wormhole. The structure and the shape of this shadow will be different from a black hole shadow \cite{Bronnikov:2021liv,Cardoso:2014sna,Ohgami:2015nra,Paul:2019trt}. A deep understanding of null geodesics and its corresponding surface is needed before any experiment could be carried out. There are several methods to determine the existence of wormholes, see for example \cite{DeFalco:2021klh,DeFalco:2020afv}. On the other side, a mathematical framework for describing these type of surfaces is needed.  Usually, in order to understand the behavior of LR's around  the wormhole throat  a solution of radial null geodesic equation is needed. Nevertheless, a new geometric  procedure for studying the black hole photon sphere was proposed in \cite{Qiao:2022jlu,Qiao:2022hfv,Cunha:2022nyw}. Using the geodesic curvature of a curve defined using  the optic metric, a condition for the existence of LR's was deduced. Using the geodesic curvature of a curve defined over a $2-$ dimensional Riemannian surface a condition for the existence of null circular orbits was proposed, namely the vanishing of that geodesic curvature. Similarly, a criteria for studying the stability of LR's was proposed. It uses the sign of the Gaussian curvature of the optic metric. A generalization to timelikce circular orbits  was presented in \cite{Bermudez-Cardenas:2024bfi, Bermudez-Cardenas:2025duw}, where a condition for the existence of massive particle surfaces was deduced \cite{Bogush:2023ojz,Bogush:2024fqj,Kobialko:2022uzj}. This geometric approach is intertwined  with techniques that help to construct a Riemannian metric from a spacetime \cite{Arganaraz:2021fwu}. In this article we take a similar path, we apply geometric techniques, namely intrinsic curvatures, to study LR's and TCO's in wormhole geometries, although the first steps were taken in \cite{Arganaraz:2019fup,Duenas-Vidal:2022kcx,Li:2019vhp}.\\
A wormhole spacetime can also have a photon sphere and an accretion disk \cite{Zhang:2024,Bronnikov:2021liv}. Geometrically, it means that the geodesic curvature of a curve defined by a $2$-dimensional Riemannian metric obtained by projection over surfaces of constant energy vanishes. This leads to a condition for the existence of a LR. Analyzing the asymptotic limit of this geodesic curvature we are able to determine the existence of LR's and give a clue about its number. A similar procedure can be carried out for  TCO's. The geodesic curvature has to vanish but now it has to be constructed using the Jacobi metric, a $2$-dimensional Riemannian metric obtained by projecting over surfaces of constant energy. Although we perform the analysis maintaining the most general wormhole metric, it will be difficult to say something without specifying some data about the metric, specially near the throat. We focus in static spherically symmetric asymptotically flat wormholes, although our analysis can be extended to other types of wormholes.
In section \ref{ss:1} we introduce the formalism that we are going to use. We show how to obtain a Riemannian metric from a Lorentzian spacetime and we present an exact expression of the Gaussian and geodesic curvatures of that metric. In section \ref{ss:2} we apply these results to wormhole metrics. We provide a detailed study of the Morris-Thorne wormhole family of metrics. We analyze the existence of LR's and TCO's and deduce the master equation that determines the conditions for the existence of massive particle surfaces. In section \ref{ss:4} we study the Damour-Solodukhin wormhole. We study the existence of LR's and TCO's and derive the master equation that needs to be satisfied in order to get a massive particle surface. In section \ref{ss:5} we analyze the Ellis-Bronnikov wormhole.  Finally, we present the discussion section.

\section{Light rings and timelike circular orbits from curvature}\label{ss:1} 

Recently, the interest in describing the geometrical properties of the black hole shadows has lead to the development of new geometrical techniques. Instead of studying the geodesics followed by photons/particles the new tools exploit the geometric properties, usually using Riemannian submanifold theory, of the hypersurfaces formed with these photons/particles trajectories. The manifold constructed with the null trajectories is known as photon surface, and for static spherically symmetric black holes it is a totally umbilic surface\footnote{The total umbilicity property is determined by the proportionality of the first and second fundamental forms} \cite{Claudel:2000yi,Senovilla:2011np}. A similar concept was formalized using the trajectories followed by massive particles, the hypersurface built with the worldlines of particle with mass is called massive particle surface (MPS). The MPS's have a partial umbilicty property \cite{Bogush:2023ojz,Bogush:2024fqj}. Recently, it was shown that the partial umbilicity property of the MPS's can bee seen as total umbilicity but in the Jacobi metric, a Riemannian metric obtained by projecting the spacetime metric over surfaces of constant energy. Furthermore, a simple method, based on the geodesic curvature of  a curve calculated using the Jacobi metric, for determining if a hypersurface is a MSP   was developed in \cite{Bermudez-Cardenas:2024bfi,Bermudez-Cardenas:2025duw}. \\
The optic metric, a $2$-dimensional Riemannian metric, can be used to determine the existence of light rings \cite{Qiao:2022jlu,Qiao:2022hfv,Cunha:2022nyw}. A generalization  to timelike circular orbits existence can be worked by using the Jacobi metric, a metric obtained by projecting over the directions of the Killing vectors of the spacetime metric \cite{
Bermudez-Cardenas:2024bfi,Bermudez-Cardenas:2025duw,Cunha:2022nyw}. We are going to use these formalism to study the existence of light rings (LR's) and (TCO's) in wormhole metrics of the type: 
\begin{equation}\label{metricg:1}
ds^2=-f(r)dt^2+\frac{1}{g(r)}dr^2+h(r)(d\theta^2+\sin^{2}\theta d\phi^2).
\end{equation}

We impose asymptotic flatness by setting:

\begin{equation}\label{alimit:1}
\lim_{r\rightarrow \infty}f(r)=1,\,\,\,\,\,\lim_{r\rightarrow \infty}g(r)=1,\,\,\,\,\,\lim_{r\rightarrow \infty}h(r)=r^2.
\end{equation}

The Jacobi metric obtained from \eqref{metricg:1} is given by \cite{Bermudez-Cardenas:2024bfi}

\begin{equation}\label{jm:1}
J_{ij}dx^{i}dx^{j}=F(r)\left[\frac{dr^2}{g(r)}+h(r)(d\theta^2+\sin^2(\theta)d\phi^2)\right],
\end{equation}
where
\begin{equation}\label{Fjm:1}
F(r)=\frac{E^2-m^2f(r)}{f(r)}
\end{equation}

The Jacobi metric defined in \eqref{jm:1} was obtained by projecting the metric  \eqref{metricg:1} in the direction of the Killing vector $\partial_{t}$, in other words, over surfaces of constant energy. It turns out that the Riemannian metric \eqref{jm:1} encodes the properties of the photon surfaces and of MPS's. We are going to use the geodesic curvature calculated with \eqref{jm:1} for determining the existence of LR's and TCO's. We also study the conditions for the existence of photon and massive particle surfaces. 

\subsection{Gaussian and geodesic curvatures}
We are going to consider the intersection of the metric \eqref{jm:1} with the surface $\theta=\frac{\pi}{2}$, then  the metric \eqref{jm:1} is going to be a $2-$ dimensional Riemannian metric.  The Gaussian curvature calculated using  this $2-$dimensional metric  is given by \cite{Bermudez-Cardenas:2024bfi}:

\begin{eqnarray}\label{Kgen:2}
\mathcal{K}&=&-\frac{E^2}{\left(E^2-m^2f\right)^2}\left[g' W(r)+\frac{g}{2}\Omega(r)\right]
\end{eqnarray}

\begin{eqnarray}
W(r)&=&\frac{h'f-hf'-\frac{m^{2}}{\mathcal{E}^2}f^2h'}{4h}\\
\Omega(r)&=&f'\left(\frac{2f'}{f}-\frac{h'}{2h}\right)-f''-\left(\frac{(h')^2}{2h^2}-\frac{h''}{h}\right)f-\frac{(f')^2}{\left(1-\frac{m^2}{E^2}f\right)f}\nonumber\\
& &+\frac{m^2}{E^2}\left(\frac{(h')^2}{2h^2}-\frac{h''}{h}\right)f
\end{eqnarray}

For asymptotically flat spacetimes $\lim_{r\rightarrow \infty}\mathcal{K}=0$, therefore, the Jacobi metric is an asymptotically flat Riemannian metric. Setting $m=0$ in expression \eqref{Kgen:2} we will be able to apply the criteria for the existence and stability of LR's presented in \cite{Qiao:2022jlu,Qiao:2022hfv}. We only need to specify the components of the metric \eqref{metricg:1}, we are going to work with the components of a wormhole metric.\\
Similarly, we can calculate a geodesic curvature using the metric \eqref{jm:1}. This geodesic curvature is calculated for circular trajectories defined over a $2-$dimensional surfaces. If it vanishes then the curve is a LR or a TCO depending if it is null or timelike respectively. Therefore, this geodesic curvature is going to be used to determine the existence of light rings and TCO's in wormhole spacetimes. We will use the criteria developed in \cite{Bermudez-Cardenas:2024bfi} to find an expression for the energy per unit mass, this expression is going to help us to determine the existence of a MPS's and to calculate the radius of the ISCO (Innermost Stable circular orbit) of the wormhole.\\
The geodesic curvature obtained from \eqref{jm:1} is given by
\begin{equation}\label{geo:1}
\kappa_{g}=\frac{1}{2}\sqrt{\frac{g}{F}}\left(\frac{F'}{F}+\frac{h'}{h}\right),
\end{equation}
where $F(r)$ is given by \eqref{Fjm:1} with $g(r)$ and $h(r)$ being the components of the metric  \eqref{metricg:1}. Thus, using equation  \eqref{Fjm:1}  the expression  \eqref{geo:1} can be written
\begin{equation}\label{geo:2}
\kappa_{g}=\frac{\sqrt{g(r)}}{\mathcal{E}\left(1-\frac{m^2}{E^2}f(r)\right)^{3/2}\sqrt{f(r)}}\left(\frac{h'f-hf'-\frac{m^2}{E^2}h'f^2}{2h}\right).
\end{equation}

A simple expression that relates the existence of the LR's/TCO's   was found  by equating the geodesic curvature \eqref{geo:2} to zero\cite{Bermudez-Cardenas:2024bfi}.
\begin{equation}\label{cond:1}
\partial_{r}(F h)=h'f-hf'-\frac{m^2}{E^2}h'f^2=0.
\end{equation} 

When $m=0$ the expression \eqref{cond:1} transforms to $h'f-hf'=0$, which is a well known expression for the existence of  LR's \cite{Claudel:2000yi}. We are going to use \eqref{cond:1} to study the existence of TCO's in wormhole geometries.

\section{LR's and TCO's of static, spherically symmetric wormholes}\label{ss:2}

The existence of LR in black hole espacetimes have been studied extensively. In  \cite{DiFilippo:2024ddg} using only geometric properties, a theorem for the stability of  the inner LR of an UCO (Ultra-compact object) was presented. The existence and stability of LR's in UCO's has been also analyzed in \cite{Ghosh:2021txu} see also \cite{Cunha:2020azh,Cunha:2017qtt,Guo:2020qwk}. Using a different approach, the existence of LR's in wormhole spacetimes was studied in \cite{Xavier:2024iwr}. In this section we want to study the existence of LR's for wormhole spacetimes using a Riemannian metric, the so called Jacobi metric, defined in \eqref{ss:2}. The geometric approach was presented in \cite{Cunha:2022nyw,Bermudez-Cardenas:2024bfi}, see also \cite{Qiao:2022jlu,Qiao:2022hfv}. We start by analyzing what happens with the geodesic curvature \eqref{geo:2} at infinity. Replacing  conditions \eqref{alimit:1} together with $m=0$ in \eqref{geo:1} we get
\begin{equation}\label{limkglr}
\lim_{r\rightarrow \pm \infty}\kappa_{g}=\frac{1}{E}\frac{h'(r)}{2h(r)}\sim \pm \frac{1}{r}.
\end{equation}
The function $\kappa_g$ goes to zero at $+\infty$, one side of the wormhole,  and also goes to zero at $-\infty$, at the other side of the wormhole. The interesting behavior  is that on one side $r>0$ the geodesic curvature goes to zero from positive values and on the other side $r<0$ the geodesic curvature goes to infinity from negative values, then by the intermediate value theorem we have that there is at least one point where the geodesic curvature vanishes. In other words, there exist at least one LR, a similar result was obtained in \cite{Xavier:2024iwr}. We have more, note that geodesic equation vanishes at $g(r)=0$, it implies that there is a LR where the throat is located.
If we know the near throat behavior of \eqref{geo:2} we can study in detail what happens with LR's/TCO's at both sides of the throat.
The existence of TCO's can be also studied following the method used for LR's. Thus, taking the asymptotic limit  of the geodesic curvature \eqref{geo:2} and maintaining $m\neq 0$ we get
\begin{equation}\label{limkgtco}
\lim_{r\rightarrow \pm \infty}=\frac{1}{E}\frac{h'(r)}{2h(r)}\sim \pm \frac{1}{Er}.
\end{equation}
The limit in expression \eqref{limkgtco} is the same as in \eqref{limkglr}, it has the same implications for TCO's, namely there should be at least one TCO. This result represent a generalization of the one obtained in \cite{Xavier:2024iwr}, see also \cite{Cunha:2022gde}. When $m=0$  we just recover the already known results for LR's. It shows that our formalism is so powerful that we can carry out the analysis of LR's and TCO's at the same time. If we want to extract more information we need to specify the wormhole functions. In the following we are going to study different type of wormholes and we will show how the formalism described in this section can be applied to these horizonless asymptotically flat spacetimes.

\section{Morris-Thorne wormhole}\label{ss:3}
The Morris-Thorne wormhole metrics \cite{Morris:1988cz} constitutes a family of metrics which are defined by two functions: the redshift function $\Phi(r)$ and the shape function $b(r)$. The metric can be written
\begin{equation}\label{MTworm}
ds^2=-e^{2\Phi(r)}dt^2+\frac{dr^2}{1-\frac{b(r)}{r}}+r^2\left(d\theta^2+\sin^2{\theta d\phi^2}\right).
\end{equation}
In order to ensure the absence of horizons it is required that $g_{tt}=e^{-2\Phi}\neq 0$ and therefore $\Phi(r)$ must be finite everywhere. The shape function $b(r)$ has a minimum at the throat of the wormhole,  which is located in $r_o$, and therefore $b(r_o)=r_o$. The Jacobi metric associated to the metric \eqref{MTworm} is
\begin{equation}\label{Jacobimt}
J_{ij}dx^{i}dx^{j}=\left(\frac{E^2-m^2 e^{2\Phi(r)}}{e^{2\Phi(r)}}\right)\left(\frac{dr^2}{\left(1-\frac{b(r)}{r}\right)}+r^2\left(d\theta^2+\sin^2{\theta d\phi^2}\right)\right).
\end{equation}

Comparing \eqref{Fjm:1} with \eqref{Jacobimt}  the $F(r)$ function can be identified:
\begin{equation}\label{Fgen:1}
F(r)=\left(\frac{E^2-m^2 e^{2\Phi(r)}}{e^{2\Phi(r)}}\right).
\end{equation}

Replacing \eqref{MTworm} in \eqref{Kgen:2} we obtain an expression for the Gaussian curvature 
\begin{eqnarray}\label{Gc:1}
\mathcal{K}&=&-\frac{e^{2\Phi}}{2r^3(E^2+e^{2\Phi}m^2)}\left[m^4 e^{4\Phi}(b-rb')
\right.\nonumber\\
& &
\left.
-e^{2\Phi}E^2 m^2\left(2r^2((1-2r\Phi')\Phi'+r\Phi'')+b'r(2-r\Phi')-b\left(2+r(\Phi'(1-4r\Phi')+2r\Phi'')\right)\right)
\right.\nonumber\\
& &
\left.
E^4(-2r^2(\Phi'+r\Phi'')+b'r(r\Phi'-1)+b(1+r(\Phi'+2r\Phi'')))\right].
\end{eqnarray}

It is easy to see that when we set the redshift function $\Phi(r)=0$ the Gaussian curvature transforms to:
\begin{eqnarray}\label{mtK}
\mathcal{K}=\frac{b'r-b}{(E^2-m^2)r^3}.
\end{eqnarray}

which is in accordance with the result obtained in \cite{Arganaraz:2019fup}. The sign of \eqref{mtK} is determined by the numerator which turns out to be the flare-out condition of the wormhole, therefore it has a definite sign, therefore the LR located at the throat of the wormhole is unstable. Note that here we do not used the geodesic curvature because of the flare-out condition of the wormhole, which states that for a throat of the Morris-Throne wormhole it has to be satisfied that $b'r-b<0$.\\
Let us show how to use the geodesic curvature of the Jacobi metric \eqref{Jacobimt} to determine the existence of LR's and TCO's. Thus, the geodesic curvature of the Jacobi metric \eqref{Jacobimt} is

\begin{equation}\label{mtgeo}
\kappa_{g}=\frac{e^{2\Phi}\left(1-\frac{b}{r}\right)^{1/2}}{r(E^2+e^{2\phi}m^2)^{3/2}}\left(E^2+e^{2\Phi}m^2-E^2r\Phi'\right)
\end{equation}

A TCO has to satisfy $\kappa_{g}=0$. Therefore, from the numerator of \eqref{mtgeo} we obtain two factors, the first one leads to the condition
\begin{equation}
b(r_o)=r_o,
\end{equation}
which tell us that a TCO is located at the throat. This result was found in \cite{Arganaraz:2019fup} for the case $\Phi=0$. It shows that no matter the form of $\Phi(r)$, there is always a LR at the throat. We have more, from  the second factor in \eqref{mtgeo} we get
\begin{equation}\label{mtecond}
\frac{E^2}{m^2}=\frac{e^{-2\Phi} }{(1-r\Phi')},
\end{equation}
which shows that the energy per unit mass of the orbit only depends on the redshift function. There is not an analogous result for the case $\Phi=0$ because the condition \eqref{mtecond} implies that $E^2=m^2$ which is a singular point of the Gaussian curvature. When $\Phi\neq 0$ the singular point of the Jacobi metric moves to where $e^{-2\Phi}E^2-m^2=0$. Note that $1-r\Phi'=0$ is a singular point of the Gaussian curvature and the expression \eqref{mtecond} becomes also singular. Therefore, if we want to build a wormhole metric that has a photon sphere, redshift functions of the form $\Phi\neq ln(r)$ are ruled out. The radius of a photon sphere is found by solving $1-r\Phi'=0$. The ISCO orbit can be obtained by solving the equation $dE/dr=0$, thus
\begin{equation}
-m^2\Phi'(2r\Phi'(r)-r\Phi''-3)=0.
\end{equation}

Circular geodesics exist only if $1-r\Phi'>0$. We want to see if there are more points where $\kappa_g=0$. In order to see it we need to analyze the near throat and near infinity limit. \\
In the metric \eqref{MTworm} the redshift function $\Phi(r)$ and the shape function $b(r)$ have asymptotic properties, the redshift function should behave at spatial infinity in such a way that $e^{2\Phi}\rightarrow 1$, therefore $\Phi\rightarrow 0$. Similarly, the shape function  has to satisfy $\frac{b(r)}{r}\rightarrow 0$ at spatial infinity. Using these asymptotic properties we can see that the Gaussian curvature \eqref{Gc:1} vanishes at infinity  as expected. On the other side, the geodesic curvature \eqref{mtgeo} has two factors in the numerator, each one depending on the shape function $b(r)$ and the redshift function $\Phi$ respectively. Therefore,we have
\begin{equation}
\lim_{r\rightarrow \pm\infty}\kappa_g=\pm\frac{1}{r(E^2+m^2)^{1/2}}.
\end{equation}
from the previous limit we can conclude, as in the general case, that the Morris-Thorne wormhole metrics has at least one LR. It is direct to conclude that if $\kappa_g$ has to go from negative values at $-\infty$ towards positive values at $+\infty$,  because of the intermediate value theorem, it has at least one zero. Not only that, if $k_{g}$ has more zeros then it has to be an odd number.  Evidently, the same conclusion  can be stated for TCO's. We are showing that the theorem that affirms that the number of LR's is odd  can be extended to TCO's, an as before, one of them is located at the throat.  Let see how it works on different scenarios.

\subsection{Case I: $b(r)=b_o^{n}r^{1-n}\,\,n>2,\,\,and$  $\Phi(r)=0$}

Let us consider the shape function $b(r)=b_{o}^{n}r^{1-n}$ with $n>2$, and a redshift function $\Phi=0$,  then the geodesic curvature \eqref{geo:2} satisfies:
\begin{equation}
\lim _{r\rightarrow\pm\infty}k_{g}=\frac{1}{(E^2+m^2)^{1/2}}\left(\frac{1}{r^2}-\frac{b_o^n}{r^{n+2}}\right)^{1/2}=\pm\frac{1}{r(E^2+m^2)^{1/2}}\sim 0
\end{equation}
The result was expected. At infinity the spacetime is flat and geodesic are straight lines and if $m=0$ the result is the same. 
For metric \eqref{MTworm} the throat is located at $r_{o}=b_{o}$, then the geodesic curvature \eqref{mtgeo} has the property:
\begin{equation}
\lim_{r\rightarrow b_{o}}k_{g}=0.
\end{equation}
It implies that when we reach the throat we are at a circular geodesic. This is in agreement with \cite{Arganaraz:2019fup}. In general, the geodesic curvature is going to vanish at the throat and at infinity, with the difference that in one side it reaches zero from positive values and on the other from negative values. The Gaussian curvature near the throat became
\begin{equation}
\lim_{r\rightarrow bo}\mathcal{K}=-nb_{o}.
\end{equation}

Due to the fact that the Gaussian curvature near the throat is negative independent of the radial coordinate we conclude that the LR located at the throat is unstable.\\
Finally, we want to see if the derivative of the geodesic curvature \eqref{Gc:1} vanishes at any point different from the throat, this could give a clue about the number of LR's/TCO's, which until now we only know that there is an odd number. Taking the derivative of $\kappa_g$ and equating it to zero we obtain
\begin{equation}
\left(E^2-\delta \right) \left(r \left((1-n) b_o^n
r^{-n}+2\right)-3 b_o^n r^{1-n}\right)=0.
\end{equation}
Hence, the derivative of the geodesic curvature vanishes when 
\begin{equation}
r_{*}=b_o\left(1+\frac{n}{2}\right)^{1/n},
\end{equation}
then 
\begin{equation}
\kappa_{g}(r_{*})=\frac{n}{2bo^2\left(1+\frac{n}{2}\right)^{n/2}}>0.
\end{equation}
From the previous equation we can conclude that the derivative of the geodesic curvature vanishes at two points. Then the Morris-Thorne wormhole with $\Phi=0$ and $b=b_o^nr^{1-n}$ has only one LR localized at the throat, and as we conclude from the sign of the Gaussian curvature , this orbit is unstable. In the following sections  we will see that this behavior is very common.

\subsection{Case II: $\frac{b(r)}{r}=\frac{8r_{o}}{3r}-\frac{5r_{o}^2}{3r^2}.\,\,and $  $e^{2\Phi}=1-\frac{8r_o}{3r}+\frac{15 r_o^2}{8r^2}$}

As we have shown in previous sections, the curvatures of the Jacobi metric carry information about the photon spheres and massive particle surfaces. Here we consider a wormhole such that \cite{Shaikh:2018oul}
\begin{equation}\label{worm:1}
e^{2\Phi}=1-\frac{8r_o}{3r}+\frac{15 r_o^2}{8r^2},\,\,\,\,\,1-\frac{b(r)}{r}=1-\frac{8r_{o}}{3r}+\frac{5r_{o}^2}{3r^2}.
\end{equation}

The wormhole defined by \eqref{worm:1} was built to have an effective photon sphere at its throat and a photon sphere outside its throat. Replacing \eqref{worm:1} in \eqref{mtgeo} and enforcing the condition for the existence of circular geodesics $\kappa_g =0$ we obtain
\begin{equation}
r_1=\frac{3r_o}{2},\,\,\,\,\,r_2=\frac{5r_o}{2}.
\end{equation}
In both $r_1$ and $r_2$ the sign of the Gaussian curvature is negative and therefore, the the orbits are unstable.

\section{Damour-Solodukhin wormhole}\label{ss:4}

One of the simplest way of building a wormhole metric is by modifying the metric of a black hole in such a way that the horizon is removed. Indeed, the Schwarzschild metric can be modified by adding a dimensionless parameter $\lambda^2$ such that the term $g_{tt}$ does not diverge anymore, then, in $r=2M$ instead of a horizon there is a throat that joins two asymptotically flat regions that are isometric. The metric of such a wormhole can be written as \cite{Damour:2007ap} :

\begin{equation}\label{darmourw}
d s^2=-(g(r)+\lambda^2)dt^2+\frac{dr^2}{g(r)}+r^2(d\theta^2+\sin^{2}(\theta)d\phi^2).
\end{equation}
The study of null and timelike orbits has been carried out in \cite{Tsukamoto:2024pid,Tsukamoto:2019ihj,Tsukamoto:2020uay}. The corresponding Jacobi metric of \eqref{darmourw} is 
\begin{equation}\label{jmdarmour}
J_{ij}dx^idx^j=\left(\frac{E^2}{g(r)+\lambda^2}-m^2\right)\left(\frac{dr^2}{g(r)}+r^2(d\theta^2+\sin^2(\theta)d\phi^2)\right).
\end{equation}
replacing the metric \eqref{darmourw} in the expression of the Gaussian curvature  \eqref{Kgen:2}  we get:

\begin{equation}\label{Gds:1}
\mathcal{K}=\frac{G(r)}{4 r \left(g(r)+\lambda
   ^2\right) \left(m^2 \left(g(r)+\lambda ^2\right)-E^2\right)^3}
\end{equation}
where
\begin{eqnarray}
G(r)&=&2 E^2 r g(r) \left(g+\lambda ^2\right) g'' \left(m^2 \left(g+\lambda ^2\right)-E^2\right)\nonumber\\
& &+2 \left(g+\lambda ^2\right) g' \left(\lambda ^2 \left(E^2-\lambda ^2 m^2\right)^2-m^2 \left(-g^2 \left(3 \lambda ^2 m^2-E^2\right)
\right. \right.\nonumber\\
& & 
\left.\left.
+3 E^2 g \left(\lambda ^2+3 \lambda ^4 m^2\right)-m^2 g^3\right)\right)\nonumber\\
& &+E^2 r g'^2 \left(E^2 g+E^2 \left(-\lambda ^2\right)+m^2 \left(-2 \lambda ^2 g-3 g^2+\lambda ^4\right)\right)
\end{eqnarray}
Similarly, the geodesic curvature \eqref{geo:1} now becomes:

\begin{equation}
k_{g}=-\frac{g(r)^{1/2} \left(E^2 r g'+2 \left(g+\lambda ^2\right) \left(m^2 \left(g+\lambda ^2\right)-E^2\right)\right)}{2 r \left(g+\lambda ^2\right)^{1/2} \left(m^2 \left(g+\lambda ^2\right)-E^2\right)^{3/2}}
\end{equation}

By looking at points where $\kappa_g$ vanishes we find circular geodesics but also we derive the master equation for the existence of massive particle surfaces:
\begin{equation}\label{enD:1}
\frac{E^2}{m^2}=\frac{2(\lambda^2+g)^2}{2(\lambda^2+g)-rg'}.
\end{equation}
The right side of the previous expression is positive only if $2(\lambda^2+g)-rg'>0$. Moreover, the radius of a photon sphere is obtained by solving $2(\lambda^2+g)-rg'=0$. \\
In order to determine the existence of LR's and TCO's we need to know what happens with the curvatures near the throat of the wormhole and at spatial infinity. In order to do so, we take \cite{Damour:2007ap}  $g(r)=1-\frac{2M}{r}$, then the Gaussian curvature \eqref{Gds:1} becomes:

\begin{equation}
\mathcal{K}_{DS}=\frac{G(r)}{r^3 \left(-2 M+r(\lambda ^2 +1)\right) \left(m^2
   \left(r(\lambda ^2+1)-2 M\right)-E^2 r\right)^3},
\end{equation}
where
\begin{eqnarray}
G(r)&=&M \left(E^4 r^2 \left(6 M^2-7 \left(\lambda ^2+1\right) M r+\left(\lambda ^4+3 \lambda ^2+2\right) r^2\right)
\right.\nonumber\\
& &
\left.
+m^2
   \left(\left(r(\lambda ^2 +1)-2 M+\right)^4-E^2 r \left(-12 M^3+4 \left(7 \lambda ^2+6\right) M^2 r
\right.\right.\right.\nonumber\\
& &
\left.\left.\left.
-15 \left(\lambda ^2+1\right)^2 M
   r^2+\left(\lambda ^2+1\right)^2 \left(2 \lambda ^2+3\right) r^3\right)\right)\right).
\end{eqnarray}
The previous expression satisfies
\begin{equation}
\lim_{r\rightarrow \infty}\mathcal{K}_{DS}=0,
\end{equation}
therefore at spatial infinity the Gaussian curvature goes to zero, then the Jacobi metric of \eqref{jmdarmour} is asymptotically flat. On the other side, near the throat the Gaussian curvature behaves
\begin{equation}
\lim_{r\rightarrow r_h}\mathcal{K}_{DS}=\frac{E^2( 1-2\lambda ^2)+2 \lambda ^4 m^2}{16 M^2 \left(E^2-\lambda ^2 m^2\right)^2}.
\end{equation}
When $m=0$ the limit behaves as:
\begin{equation}
\lim_{r\rightarrow r_h}\mathcal{K}_{DS}=\frac{E^2( 1-2\lambda ^2)}{16 M^2 E^4}.
\end{equation}
The geodesic curvature \eqref{geo:2} becomes 
\begin{equation}
\kappa_{g}=\frac{P(r)}{r^{3/2} (r(\lambda ^2+1)^{1/2} -2 M+)\left(E^2 r+m^2 \left(2
   M-\left(\lambda ^2+1\right) r\right)\right)^2}
\end{equation}
where
\begin{eqnarray}
P(r)&=&(2 M-r)^{1/2} \left(m^2 \left(r(\lambda ^2 +1)-2 M\right)-E^2 r\right)^{1/2} \left(m^2 \left(r(\lambda ^2+1)-2 M\right)^2
\right.\nonumber\\
& &
\left.
-E^2 r \left(r(\lambda ^2 +1)-3 M\right)\right)
\end{eqnarray}

It is evident that the geodesic curvature vanishes when $r=2M$, this defines a circular geodesic, for null and timelike trajectories. For $m=0$ the Gaussian curvature at $r=2M$ is going to be positive if $\lambda^2<1/2$, and therefore the LR is going to be stable, otherwise it is going to be unstable.

The geodesic curvature  $\kappa_{g}$ has two more roots \footnote{There is another root at $r=0$ but it is a singularity which is beyond the value of $r$ at the throat, therefore we will not consider it.}
\begin{equation}
r_{t}=\frac{M \left(4 \left(\lambda ^2+1\right) m^2\pm E \left(\sqrt{9 E^2-8 \left(\lambda ^2+1\right) m^2}\mp 3
   E\right)\right)}{2 \left(\lambda ^2+1\right) \left(\left(\lambda ^2+1\right) m^2-E^2\right)}.
\end{equation}
When $m=0$ the Gaussian curvature evaluated at $r_t$ can be written
\begin{equation}
\mathcal{K_{DS}}({r_t})=\frac{2 \left(\lambda ^2+1\right)^3 \left(2 \lambda ^2-1\right)}{52 E^2  M^2},
\end{equation}
which is going to be positive if $\lambda^2>1/2$, otherwise it will be negative, and therefore the LR is going to be unstable or stable respectively.\\
We were able to find the LR's and in some cases the  TCO's. In the case of wormhole given by the metric \eqref{darmourw} the results can be inferred only by the properties of $g(r)$. The wormhole has three LR's, one at the throat. Finally, from the condition $\kappa_g=0$ we can also obtain
\begin{equation}\label{enD:2}
\frac{E^2}{m^2}=\frac{(2M-r(1+\lambda^2))^2}{r(3M -r(1+\lambda^2))}.
\end{equation}
The existence of a massive particle surfaces is guaranteed. Setting $\lambda=0$ in \eqref{enD:1} we recover the result for the Schwarzschild black hole.
The radius of the photon sphere satisfies $r(3M -r(1+\lambda^2))=0$, hence
\begin{equation}
r_{ph}=\frac{3M}{1+\lambda^2}.
\end{equation}
Using $dE/dr=0$ we obtain the radius of the innermost stable circular orbit
\begin{equation}
r_{ISCO}=\frac{6M}{1+\lambda^2}
\end{equation}

\section{Ellis-Bronnikov wormhole}\label{ss:5}
One of the most known wormholes is the Ellis metric \cite{Ellis:1973yv}. This metric is built as a modification of flat spacetime  by adding a constant to the radial coordinate, it can be written as
\begin{equation}\label{ell:1}
ds^2=-dt^2+dr^2+(r^2+a^2)(d\theta^2+\sin^2(\theta)d\phi^2).
\end{equation}
The added parameter $a$ constitutes the size of the wormhole's throat, which is located at $r=0$. This  wormhole is an asymptotically flat spacetime. A generalization can be done by including radial functions $u(r)$ in the metric \eqref{ell:1} in the following way \cite{Huang:2023yqd,Ishkaeva:2023xny,Yazadjiev:2017twg}:
\begin{equation}\label{ellb:1}
ds^2=-e^{2u(r)}dt^2+e^{-2u(r)}\left(dr^2+(r^2+a^2)d\Omega^2\right),
\end{equation}
where $d\Omega^2=d\theta^2+\sin^2(\theta)d\phi^2$, $M$ is the associated mass and $a$ is an arbitrary parameter. The function $u(r)$ is given by
\begin{equation}
u(r)=\frac{M}{a}\left(\arctan\left(\frac{r}{a}\right)-\frac{\pi}{2}\right).
\end{equation}
In the asymptotic limit $r\rightarrow \pm\infty$ the metric coefficients behave
\begin{eqnarray}
e^{2u(r)}\bigg|_{r\rightarrow +\infty}&=&\left(1-\frac{2M}{r}\right)+\mathcal{O} \left(\frac{1}{r^2}\right)\\
e^{2u(r)}\bigg|_{r\rightarrow -\infty}&=&e^{\frac{2\pi M}{a}}\left(1+\frac{2M}{|r|}\right)+\mathcal{O} \left(\frac{1}{|r|^2}\right).
\end{eqnarray}
It is clear that the metric above has two asymptotically flat regions connected at the throat located at $r_{th}=M$. When $M=0$ we recover the Ellis metric \eqref{ell:1}. Using \eqref{jm:1} we can calculate the Jacobi metric associated to the  Ellis-Bronnikov metric \eqref{ellb:1}:
\begin{equation}\label{jmeb:1}
J_{ij}dx^{i}dx^{j}=\left(E^2-m^2e^{{2u(r)}}\right)\left(dr^2+(r^2+a^2)d\phi^2\right).
\end{equation}

Replacing metric \eqref{ellb:1}  in the Gaussian curvature \eqref{Kgen:2} we get
\begin{eqnarray}
\mathcal{K}_{EB}=\frac{e^{4 u(r)} \Psi(r)}{\left(a^2+r^2\right)^2
   \left(E^2-m^2 e^{2 u(r)}\right)^3}
\end{eqnarray}

where
\begin{eqnarray}\label{KEB:1}
\Psi(r)&=&\left(a^2-a+r^2\right) \left(a^2+a+r^2\right)
   \left(E^2-m^2 e^{2 u(r)}\right)^2\nonumber\\
 & & +\left(a^2+r^2\right) \left[2
   E^4 \left(\left(a^2+r^2\right) u''(r)+r u'(r)\right)
   \right.\nonumber\\
   & &
   \left.
 +E^2
  m^2 e^{2 u(r)} \left(-3 \left(a^2+r^2\right) u''(r)+2
   \left(a^2+r^2\right) u'(r)^2-3 r u'(r)\right)
   \right.\nonumber\\
   & &
   \left.
   +m^4 e^{4 u(r)}
   \left(\left(a^2+r^2\right) u''(r)+r.
   u'(r)\right)\right]
\end{eqnarray}
The previous Gaussian curvature will help to determine the stability of the circular orbits. In the null case ($m=0$), expression \eqref{KEB:1} becomes
\begin{equation}
\mathcal{K}_{EB}\big |_{m=0}=\frac{\left(a^4-a^2+2a^2 r^2-2 M r+r^4\right) e^{\frac{2 M \left(2
   \tan ^{-1}\left(\frac{r}{a}\right)+\pi \right)}{a}}}{E^2
   \left(a^2+r^2\right)^2}.
\end{equation} 
The sign of $\mathcal{K}_{EB}$ is determined by the polynomial $p(x)=a^4-a^2+2a^2 r^2-2 M r+r^4$. On the other side, the existence of circular geodesics is determined by the geodesic curvature. Thus, the geodesic curvature of a circular orbit in the surface where \eqref{jmeb:1} is defined is given by
\begin{equation}\label{gkeb:1}
\kappa_{EB}=\frac{1}{2(E^2-m^2e^{2u(e)})^{3/2}(r^2+a^2)}\left(2r E^2-2m^2e^{2u(r)}\left(r+u'(r)(r^2+a^2)\right)\right).
\end{equation}
When geodesic curvature vanishes we say that the curve is a circular geodesic. It vanishes at $r_{*}$ such that
\begin{equation}
E^2 (2 M-r_{*})-m^2 (M-r_{*}) e^{\frac{M \left(2 \arctan
   \left(\frac{r_{*}}{a}\right)+\pi \right)}{a}}=0.
\end{equation}
The previous equation cannot be solved analytically. The null case leads to $r_{*}=2M$, therefore, there is a LR outside the throat.\\
Imposing  that the expression in \eqref{jmeb:1} vanishes, then we obtain
\begin{eqnarray}
\frac{E^2}{m^2}&=&\frac{e^{2 u(r)} \left(a^2 u'(r)+r^2 u'(r)-r\right)}{2 (a^2 + r^2) u'(r)-r}\\
&=&\frac{(M-r) e^{\frac{M \left(2 \tan ^{-1}\left(\frac{r}{a}\right)+\pi \right)}{a}}}{2 M-r}
\end{eqnarray}
Which is the master equation for the existence of massive particle surfaces. Due to the fact that $r>2M$ the previous expression is always positive and therefore there is always a MPS. When the denominator is set to zero we obtain the radius of the photon sphere
\begin{equation}
r_{ph}=2M.
\end{equation} 
As before, by using $dE/dr=0$ we obtain the two roots, the innermost stable is going to be the smallest one
\begin{equation}
r_{ISCO}=3M-\sqrt{a^2+5M^2}
\end{equation}
Note that the value $r_{ISCO}$ has to be bigger than the throat.\\
It is direct to see that Gaussian and geodesic curvatures vanish at infinity, this is related with the asymptotic flatness of the Ellis-Bronnikov wormholes. On the other side, when we approach the throat at $r_{th}=M$ the Gaussian curvature for null orbits behaves as
\begin{equation}
\lim_{r\rightarrow M}\mathcal{K}\big|_{m=0}=\frac{\left(a^4+a^2 \left(2 M^2-1\right)+M^4-2 M^2\right) e^{2u(r)}}{E^2\left(a^2+M^2\right)^2},
\end{equation}
hence for $M>1$ we have that  $\lim_{r\rightarrow M}\mathcal{K}\big|_{m=0}>0$, then if there is a LR or a TCO at the throat it is stable. Indeed, since the LR is located at $2M$ we can evaluate the Gaussian curvature at the LR which gives
\begin{equation}
\mathcal{K}(2M)\big|_{m=0}=\frac{\left(a^2+4 M^2-1\right) e^{\frac{2 M \left(2 \tan ^{-1}\left(\frac{2
   M}{a}\right)+\pi \right)}{a}}}{\text{EE}^2 \left(a^2+4 M^2\right)}.
\end{equation}
The Gaussian curvature $\mathcal{K}(2M)\big|_{m=0}$ is going to be positive if $a^2+4M^2>1$ and therefore the LR is going to be stable.\\
Let see what happens at the throat with the geodesic curvature. The geodesic curvature near the throat becomes
\begin{equation}
\lim_{r\rightarrow M}\kappa_{EB}=\frac{E^2Me^{u(r)}}{(a^2+M^2)(E^2-e^{u(r)} m^2)^{3/2}},
\end{equation}

which is going to be positive whenever $E^2>e^{u(r)}m^2$. In the null case the condition is trivially satisfied and therefore $\kappa_{EB}\big|_{m=0}>0$. It implies $\kappa_{EB}\big|_{m=0}$ has to descend towards zero.

\section{Discussion and final remarks}\label{ss:6}

We have studied the existence of LR's and TCO's in wormhole spacetimes. We have used a purely geometric approach based on intrinsic curvatures of a Riemannian metric. Although the optic and the Jacobi metric have been used almost never in wormhole spacetimes, we have shown that it can be done and similar conclusions to the case of black holes or UCO's can be extracted. Contrary to what happens in black holes cases, there are two asymptotically flat regions connected by a zone where the radial coordinate cannot be smaller. Therefore, instead of looking at the behavior of geodesic and Gaussian curvatures at a horizon, we have to look for the asymptotic limit at both sides $\pm \infty$.   We have shown that the theorem claiming that there should be an odd number of LR can be extended to TCO's by using the Jacobi metric. In certain types of wormholes the possible number of TCO's has to be odd also.\\
Using the geodesic curvature of the Jacobi metric we have been able to determine the existence of LR's and TCO's in  spherically symmetric, asymptotically flat wormholes. The points where  the geodesic curvature vanishes are identified with LR's (when $m=0$) or TCO's (when $m\neq 0$). For a general metric, a theorem about the existence of LR's has been proposed in \cite{Xavier:2024iwr}. We have extended these previous results to TCO's and we have also studied, whenever possible, the stability of these TCO's. A condition for the existence of massive particle surfaces is also derived. We have shown that the results found for black holes can be extended almost directly to wormholes. In some cases, such as the Damour-Solodukhin wormhole, a LR can be located at the same place where a TCO's is located. The geodesic curvature can have many zeros and in principle they can be calculated, although in some cases it has to be done numerically. Instead of a direct calculation  that can be really cumbersome, the geodesic curvature of the Jacobi metric at least could give the number of LR's/TCO's.  In the case of the Morris-Thorne the method can be used to find solutions with special characteristics, for example, a solution that has a certain number of LR's. In the case of the Domour-Solodukhin wormhole we have shown that a condition for the existence of LR's is $\lambda^2>1$. In the Ellis-Bronnikov wormholes we have shown that the LR is outside the throat. In all types of wormholes that we have studied the condition for the existence of massive particle surfaces is determined. The method developed in this article can be used in different wormhole geometries, such as non-asymptotically flat spacetimes or stationary wormholes \cite{Hsieh:2024eph}, which we left for future work.  Moreover, the formalism can be applied, although with some modification, to dynamical wormholes. A general theorem regarding the existence of  TCO's in wormhole spacetimes is not known, although some steps towards this direction has been taken in this article, we left this for future work.

%%%%%%%%%%%%%%%%%%%%%%%%%%%%%%%%%%%%%%%%%%%%%%%%%%%%%%%%%%%%%%%%%%%%%%%%%%%%%%%%%%%%%%%%%%%%%%%%%%%%%%%%%%%%%%%%%%%%%%%%%%%%%%%%%%%%%%%%%%%%%%%%%%%%%%%%%%%%%%%%%%%%%%%%%%%%%%%%%%%%%%%%%%%%%%%%%%%%%%%%%%%%%%%%%%%%%%%%%%%%%%%%%%%%%%%%%%%%%%%%%%%%%%%%%%%%%%%%%%%%%%%%%%%%%%%%%%%%%%%%%%%%%%%%%%%%%%%%%%%%%%%%%%%%%%%%%%%%%%

\break

\end{document}